# ON THE POSSIBILITY OF SEPARATION OF THE "IDEAL CONDUCTIVE" AND "SUPERCONDUCTIVE" (MEISSNER STATE) PHASE TRANSITIONS.

## 2. PHYSICAL CONSEQUENCES FOLLOWING FROM THE "PARAMAGNETIC" EFFECT

### S.G. GEVORGYAN

**In the first part of this work an overview of the available data on the recently discovered in superconductors "paramagnetic" effect has been made and a possible explanation of the effect has been given. Here the consequences caused by this weakly expressed phenomenon and following from the analysis of the shape of the transition curve, are discussed, and the conclusions based on the test-data are formulated, important for true understanding of the nature of superconductivity.**

## 1. INTRODUCTION

We would like to remind that in the first part of this work [1] we presented and discussed in detail all available at the moment experimental test-data on the "paramagnetic" effect. At this, we put large attention on details of the measuring set-ups allowing us to reveal and study this unusual effect, as well as mentioned the peculiarities of samples and specified the conditions under which the effect was observed, which all are important for true understanding of the nature of this new effect. We gave also a possible explanation of the effect (not pretending, at this, on the finality of understanding of the phenomenon). Below, the possible consequences caused by this effect are given and discussed, as well as the conclusions directly following from the analysis of the experimental data are formulated.

## 2. CONSEQUENCES FOLLOWING FROM THE EXPERIMENT

In our opinion, the collected data on the "paramagnetic" (PM) effect discovered in "helium" (low-temperature) superconductors [2] and investigated then in more detail in high-$T_c$ superconductors (HTS) [3-5] (mainly, due to created by us new test-device of high resolution [6,7] measuring the superconducting (SC) properties of the matter, based on the single-layer flat pick-up coil − see first part of this work), are already enough, in order to deeply explain the physics of the SC phase transition (formation of Cooper pairs and possible changes in their behavior at further cooling of the sample − especially, at temperatures close to the transition) and make conclusions and new assumptions. In particular, this new effect (more precisely, the transition curve shape corrected by it), apparently, enables to separate from each other the "*ideal conductive*" (the state without resistance) and "*superconductive*" (the ideal diamagnetic – Meissner state) phase transitions and connect the shape of the transition curve with some normal-state physical characteristics of the material under study.

Really, the shape of the function $A/((T-T_0)/T_c)+B$ fitting the initial part of the measured SC phase transition curve shown in the inset of Fig.1 (fitting an onset of the "paramagnetic" effect), suggests that the temperature dependence of the "normal" charge-carrier density $n_n(T)$ (entering into the formula (1)

$$\delta(T) = c\Big/\sqrt{2\pi\sigma_1(T)\omega} \qquad (1)$$

for the skin depth $\delta$ [8], as a part of the normal conductance $\sigma_1(T) = [e^2 n_n(T)/m]\cdot\tau$ – see the first part of this work) cannot has the simplified form $n_n(T) = n\cdot(T/T_c)^\gamma$ assumed by the traditional "two-fluid" model [9], but, more likely, it is more complex. This dependence near the transition can be given, for example, by the relatively complex function

$$n_n(T) \cong n\cdot\{[T_c/(T_c-T_0)]^\gamma \cdot [(T-T_0)/T_c]^\gamma + n_{res}(T)/n\}, \quad T_0 < T < T_c, \qquad (2)$$

where $\gamma = 2$ (the substantiation see in the first part of this work), $n_{res}(T_c) \cong 0,$ and the senses of $T_c$ and $T_0$ is given below.

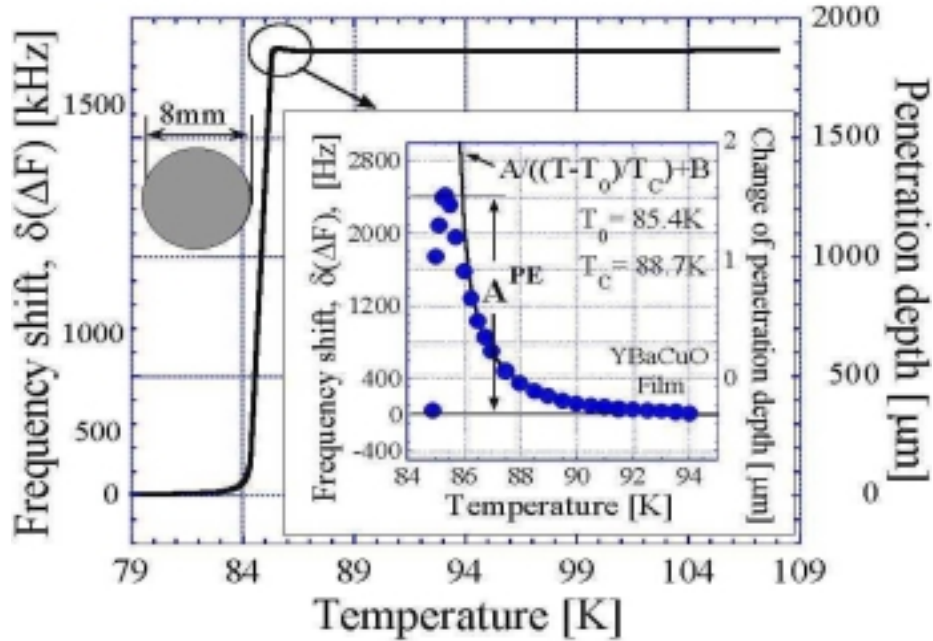

Fig.1. Superconductive phase transition curve of the tested YBaCuO film [4-5]. Inset: the enlarged view of a weakly expressed "paramagnetic" effect (PE) detected at the start of normal-to-superconductive transition, followed by the "diamagnetic" (Meissner) repulsion. $A^{PE}$ is the height of the effect and the bold line represents the fitting.

The residual density of "normal" charge carriers in this formula, $n_{res}(T)$, is a slowly rising function of the temperature, at a cooling. It is negligibly small for low-$T_c$ superconductors (LTS). However, as is known [10–13], in high-$T_c$ superconductors the residual resistance may become noticeable in many practically important applications of these materials, due to the presence of the "normal" fluid even at the absolute zero temperature ($T = 0$). In order to explain its appearance, a "three-fluid" model was proposed in [10,11] based on the concept of "non-pairing" residual normal charge carriers, $n_{res}^0 \equiv n_{res}(T = 0)$, at that, $n_{res}^0 \cong \mathrm{const}(T)$.

Temperature dependence of the form (2) means that the "normal" charge carriers (electrons) start (at the temperature $T_c \sim 88.7\mathrm{K}$ − see Fig.5 in the first part of the work) and, mainly, finish the fulfilling of their physical 'mission' in a narrow temperature region $T_c \to T_0$ (where $T_0 \sim 85.4\mathrm{K}$ is the temperature at the peak of the PM effect), and their density noticeably drops as the temperature of the specimen approaches $T_0$ (since $n_n(T_0) \cong n_{res}(T_0)$, according to Eq. (2)). Therefore, the totality of pairs with a density $n_s(T)$ in the first approximation already advances in formation **when the temperature approaches** $T_0$. At this temperature the substance, apparently, begins to possess an unusual property. Namely, ***a noticeable part of pairs*** (from all the pairs created in the substance) ***start to correlate strongly to one another***. Thus, already at these temperatures, the base of formation of the Bose condensate of Cooper pairs is lied, ***which may come out from the nulling of the total "spin" of these pairs, at*** $T_0$ ***approach*** (along with the nulling of the momentum of the pairs started for created first pairs since the temperature has been dropped down $T_c$, notifying the start of phase transition into the "*ideal conductive*" state in the substance). We suppose that the accumulation of some (minimal necessary) amount of SC pairs in a zero-"spin" (singlet) state, as the temperature $T_0$ is approached (in a medium, where pairs are present, both in the singlet and triplet states at any non-zero temperature below $T_c$), is just the reason leading to the start of the ideal diamagnetic ("*superconductive*") phase transition in a substance, as the temperature $T_0$ is approached. As this temperature is approached, the substance already makes the progress in passing into the "*ideal conductive*" state, due to enough amount of the ideally conducting Cooper pairs created in it (the particles with a zero momentum and, hence, with an infinitely large de Broglie wavelength, owing to which the pairs succeed in moving in the substance, "ignoring" defects and impurities of the crystalline structure). This is in good agreement with data presented in Fig.2, where the temperature-transition curves are shown for a YBaCuO bridge ($w \approx 0.2$ mm wide and $L \approx 4$ mm long, made of $d \approx 0.2$ μm thick film, identical to the one mentioned in Fig.1, by the chemical etching method), measured simultaneously by the standard 4-probe technique (***resistance***) and by our new test-method (by the frequency-change of the tunnel diode-based oscillator (***diamagnetic ejection***). One can see from the figure that the transition by the resistance already almost accomplishes (more than 80%) before the start of the transition by the diamagnetic properties. In other words, the ejection of the testing radio-frequency field of the coil by the sample (transition of the bridge into the Meissner state) starts when the specimen is already almost in the state without resistance. A little difference ~0.7K between the temperatures of the start of the Meissner effect and nulling of the bridge's resistance is conditioned, most probably, by worsening of the SC properties of the material along edges of the bridge during its preparation. A better coincidence might be expected, if more clean and homogeneous specimens are measured. However, for a more precise understanding of the relative positions of the resistive and diamagnetic (Meissner) phase transitions it is necessary to conduct the measurements on high-$T_c$ superconductive single crystals. But this is not so easy to do, because of very weak signals expected to detect from such small-volume (clean) physical objects. Although, according to our last experimental estimations [14], this task is quite "feasible" (from the viewpoint of higher spatial resolution of measurements) for a little improved present test-method based on a smaller (of about 1mm in diameter) flat pick-up coil.

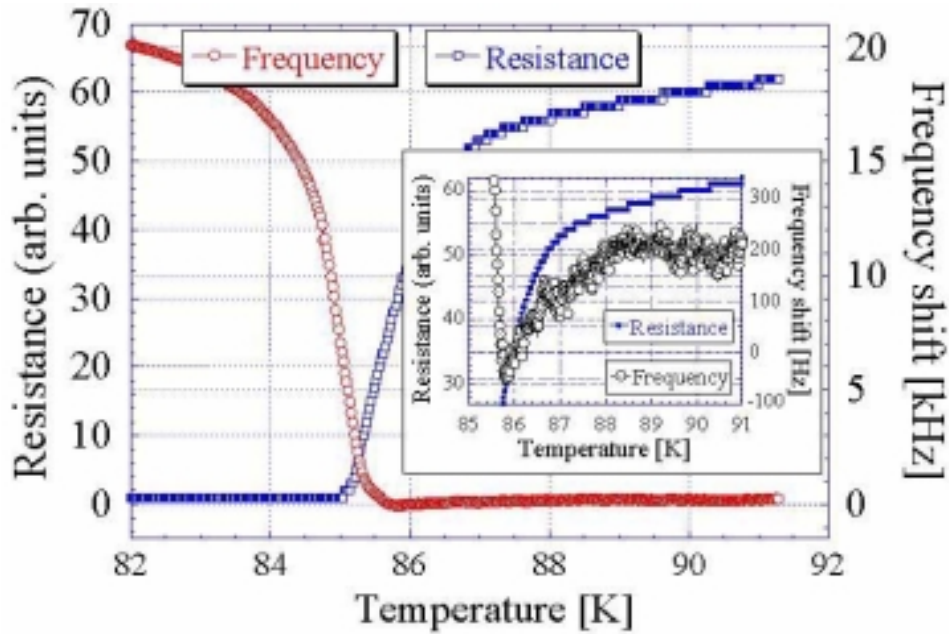

Fig.2. Superconductive phase transition curves of the YBaCuO bridge ($w \approx 0.2$ mm wide, $L \approx 4$ mm long, and $d \approx 0.2$ μm thick), made by the chemical etching of high quality film, measured simultaneously. Inset: the enlarged view of the curves close to the transition (before the diamagnetic ejection).  − standard 4-probe technique (***resistance***). o − new test-method based on a flat pick-up coil, schematically shown in Fig.1 of the first part of this work (by the frequency-shift of the tunnel diode-based oscillator − ***diamagnetic ejection***).

Further cooling of the sample below $T_0$ leads, most likely, to the changes in the behavior of the SC matter described by the commonly known BCS theory (working well in the case of "helium" (simple) superconductors [15]) on the basis of still existing in the matter (but, a few) "normal" charge carriers (decreasing by the law $n_n(T) \cong n_{res}^0 + [n_{res}(T_0) - n_{res}^0] \cdot (T/T_0)^\gamma$; $T < T_0$), and already formed in the substance large amount of the Cooper pairs with a density $n_s(T) = n - n_n(T) \cong n - n_{res}^0 - [n_{res}(T_0) - n_{res}^0] \cdot (T/T_0)^\gamma$, ($T < T_0$), where the number of the "non-paired" residual "normal" charge carriers $n_{res}^0 \neq 0$ for "nitrogen" (complex, "oxide") SC (for comparison, $n_{res}^0 \equiv 0$ for "helium" superconductors).

Stated above conclusions come out from the corrected shape of the SC phase transition curve (Fig.1) and are opposite to the commonly recognized conceptions, according to which the positions of the resistive and diamagnetic transitions practically coincide (that is, $T_0 \cong T_c$), and the number of the Cooper pairs increases at cooling, according to the simple law $n_s(T) = n \cdot [1 - (T/T_c)^\gamma]$ [9], starting with the onset temperature of the superconductive transition $T_c$ (in our notations – according to the law $n_s(T) = n \cdot [1 - (T/T_0)^\gamma]$, and starting with the temperature $T_0$).

In conclusion, we underline that just the corrected (by the new "paramagnetic" effect) shape of the SC phase transition curve of the matter (Fig.1) leads to the idea that temperatures $T_c$ (start of the PM effect − the onset point of divergence of the curves in Fig.5 of the first part of the work) and $T_0$ (peak of the PM effect) determined by it are connected with two "ideal" properties of the SC matter (and phase transitions conditioned by them). So, the first temperature ($T_c$) points out to the start of the ideal conductance (evidently, connected with the start of formation of the Cooper pairs from free electrons with opposite momenta), and the second one ($T_0$) – announces the onset of the ideal diamagnetism (with zeroing of total "spin" of the noticeable part of pairs). In addition, according to Eq. (1), initial part of the transition curve caused by the PM effect should be connected with such an important normal-state characteristics of the SC mutter as $m$ (effective mass of an electron), $\tau$ (relaxation time of charge carriers) and $n_n$ (density of "normal" charge carriers).

How much such an interpretation of physical processes determining the "real" (corrected by the "paramagnetic" effect) shape of the normal-to-superconductive phase transition curve is true, as well as, is the relation predicted by Eq.(1) between the shape of the transition curve and the normal-state physical characteristics of the matter correct, will show additional investigations in this area. The final answers on these questions may be obtained as a result of additional experimental investigations on single-crystalline HTS specimens, with the use of the measuring technique with much higher spatial resolution (for example, similar to the one used in the present study, but having the sizes of a flat pick-up coil of about 1 mm or less).

This study was partly supported by the state sources of the Republic of Armenia in frames of the research project no. 02-1359. The author is grateful to Profs. E.G.Sharoyan and Alex Gurevich (University of Wisconsin-Madison, USA) for useful discussions of experimental results and the problem as a whole, to A.A.Movsisyan, V.S.Gevorgyan, H.G.Shirinyan and A.M.Manukyan for comprehensive help during the preparation and at conducting the experiments, as well as to A.M.Mirzoyan and G.M.Manukyan for technical assistance at the final stage of the paper preparation.